\documentclass[letterpaper,12pt]{article} \usepackage{amsmath,amsfonts}

\providecommand{\e}{\mathrm{e}}
\providecommand{\bvec}[1]{{\mathbf{#1}}}

\providecommand{\partdiff}[1]{\frac{\partial}{\partial#1}}
\providecommand{\talpha}{{\tilde\alpha}}
\providecommand{\tA}{{\tilde A}}
\providecommand{\kint}{\int \frac{d^dk}{(2\pi)^d}\,}
\providecommand{\eikx}{\e^{-i\bvec{k}\cdot\bvec{x}}}
\providecommand{\tDelta}{{\tilde\Delta}}
\providecommand{\tc}{{\tilde c}}
\providecommand{\Ord}{{\mathcal{O}}}

\providecommand{\slD}{{\slash\!\!\!\!D}}
\providecommand{\slA}{{\slash\!\!\!\!A}}

\begin{document}

\title{Conformal Field Theory Correlators from Classical Field
Theory on Anti-de Sitter Space\\ II. Vector and Spinor Fields}
\author{W.~M\"uck\thanks{e-mail: wmueck@sfu.ca}\ ~and
K.~S.~Viswanathan\thanks{e-mail: kviswana@sfu.ca}\\
\small Department of Physics, Simon Fraser University, Burnaby,
B.C., V5A 1S6 Canada}
\date{\today}
\maketitle

\begin{abstract} 
We use the AdS/CFT correspondence to calculate CFT correlation 
functions of vector and spinor fields. The connection between the 
AdS and boundary fields is properly treated via a Dirichlet boundary
value problem.
\end{abstract} 
\newpage

\renewcommand{\baselinestretch}{1.2}
\normalsize

\section{Introduction}
\label{intro}
The study of conformal field theories (CFT's) in dimensions larger than
$2$ \cite{Osborn,Erdmenger} has recently been boosted by Maldacena's 
conjecture that the large $N$ limit of certain conformal field theories
in $d$ dimensions can be described by supergravity and string theory on
$d+1$ dimensional Anti-de Sitter (AdS) space \cite{Maldacena}.
Subsequently, this conjecture has been given a more precise formulation
\cite{Gubser,Witten} and it has been shown that,
in fact, any field theory on AdS$_{d+1}$ is linked to a 
conformal field theory on the AdS boundary \cite{Witten}. This
observation is entirely due to the fact that one obtains a metric on the
AdS boundary by multiplying the AdS metric with a function, which has a
single zero on the boundary in order to counteract the divergence of the 
AdS metric. However, this function is otherwise arbitrary, which imposes
the symmetries of the conformal group on the boundary metric. All one
needs then is a suitable connection between the fields on AdS$_{d+1}$
and its boundary. Schematically, this connection is given by 
\begin{equation}
\label{intro:adscft}
   Z_{AdS}[\phi_0] = \int_{\phi_0} {\mathcal{D}}\phi\, \exp(-I[\phi])
  \equiv Z_{CFT}[\phi_0] = \left\langle \exp\left(\int d^dx \,
  {\mathcal{O}} \phi_0\right) \right\rangle,
\end{equation}
where $\phi_0$ is a suitably defined boundary value of 
the AdS field $\phi$ and couples as a current to the boundary conformal 
field theory operator $\mathcal{O}$. In the classical approximation the
path integral on the l.h.s.\ is, of course, redundant. 

Field theories on AdS spaces have been the subject of research in the past
\cite{Fronsdal1,Fronsdal2,Fronsdal3,Fronsdal4,Burgess,Inami,Allen1,Allen2,Burges}.
More recently, the AdS/CFT correspondence has been investigated for 
scalar fields \cite{Volovich,Mueck1,Freedman}, gauge fields \cite{Freedman}, 
spinors \cite{Henningson}, classical gravity \cite{Liu} and type IIB string
theory \cite{Banks,Chalmers}. For a comprehensive list of recent references see
\cite{Freedman}. 

Using as representation of AdS$_{d+1}$ the upper half space $x_0>0$,
$x_i \in \mathbb{R}$, with the metric
\begin{equation}
\label{intro:metric}
  ds^2 = \frac{1}{x_0^2} dx^\mu dx^\mu,
\end{equation}
($\mu=0,1,\ldots d$), its boundary is compactified ${\mathbb{R}}^d$
(the points with $x_0=0$ and the single point $x_0=\infty$). We will frequently 
denote AdS vectors by $(x_0,\bvec{x})$ and use $x_i$ to specify the 
components of $\bvec{x}$.

The fact that the AdS metric diverges on the boundary presents a
difficulty in the AdS/CFT correspondence, which is to be met with care.
The natural solution is to calculate the AdS action on a surface
$x_0=\epsilon$ and then take the limit $\epsilon\to0$. However,
the exact connection between the AdS fields $\phi$ and
the boundary fields $\phi_0$ is subtle.
Whereas Witten \cite{Witten} stated that $\phi$ should approach $\phi_0$
times a certain power of $x_0$ as $x_0\to0$, it was soon realized
\cite{Freedman} that in certain cases, in order to satisfy Ward 
identities, one must 
formulate a proper Dirichlet boundary value problem on the surface 
$x_0=\epsilon$ and take the limit $\epsilon\to0$ at the very end.
A detailed investigation 
taking into account this subtlety has so far been done only for
scalar fields \cite{Mueck1,Freedman}. We find it therefore necessary to 
extend our previous investigation of the scalar field \cite{Mueck1} to
the vector and Dirac fields on AdS$_{d+1}$. To be general, we shall
include a mass term in the vector field action, which is considered in
Sec.~\ref{vec}. In Sec.~\ref{spin} we will give account of the
Dirac field. The minimal coupling of the Dirac 
and gauge fields is considered in Sec.~\ref{int}, and Sec.~\ref{conc}
contains conclusions.

\section{The Vector Field}
\label{vec}
The starting point is the action
\begin{gather}
\label{vec:action}
  I = \int d^{d+1}x \sqrt{g} \left( \frac14 F_{\mu\nu} F^{\mu\nu} 
  +\frac12 m^2 A_\mu A^\mu \right)\\
\intertext{with the usual relation $F_{\mu\nu} = \partial_\mu A_\nu - 
\partial_\nu A_\mu$. The equation of motion derived from \eqref{vec:action} is}
\label{vec:eqnmot}
  \nabla_\mu F^{\mu\nu} -m^2 A^\nu =0,\\
\intertext{which implies the subsidiary condition}
\label{vec:cond}
  \nabla_\mu A^\mu =0.
\end{gather}

Within our representation of Anti-de~Sitter space \eqref{intro:metric} 
one can use \eqref{vec:eqnmot} and \eqref{vec:cond} to obtain an equation 
for $A_0$, 
\begin{equation}
\label{vec:a0eqn}
  \left[ x_0^2 \partial_\mu \partial_\mu +(1-d) x_0 \partial_0
  - (m^2 - d+1) \right] A_0 =0.
\end{equation}
Introducing $\tilde m^2 = m^2-d+1$ we know from the consideration of the scalar 
field that the solution of \eqref{vec:a0eqn}, which does not diverge for 
$x_0\to\infty$, is given by
\begin{gather}
\label{vec:a0}
  A_0(x) = \kint \eikx x_0^\frac{d}{2} a_0(\bvec{k}) K_\talpha(kx_0).\\
\intertext{with}
  \talpha = \sqrt{\frac{d^2}{4} + \tilde m^2} = \sqrt{\frac{(d-2)^2}{4}
  +m^2}.
\end{gather}

It is useful to introduce fields with Lorentz indices by 
\begin{equation}
\label{vec:tadef}
  \tA_a = e^\mu_a A_\mu= x_0 A_a,
\end{equation}
where $e^\mu_a$ denotes the vielbein ($a=0,1,\ldots d$). The virtue of this 
is seen when considering the components $\tA_i$ ($i=1,2,\ldots d$), whose equation 
of motion is again obtained from \eqref{vec:eqnmot} and \eqref{vec:cond} 
and is given by
\begin{equation}
\label{vec:aieqn}
  \left[ x_0^2 \partial_\mu \partial_\mu +(1-d) x_0 \partial_0
  - \tilde m^2 \right] \tA_i = 2x_0 \partial _i \tA_0.
\end{equation}
The solution of the homogeneous part of \eqref{vec:aieqn} can be taken 
over from $A_0$ and the inhomogeneous equation is solved by making 
a good guess as to which form the solution should have. One obtains
\begin{equation}
\label{vec:ai} 
  \tA_i(x) = \kint \eikx x_0^\frac{d}{2} \left( a_i(\bvec{k})
  K_\talpha(kx_0) +i a_0(\bvec{k}) \frac{k_i}{k} x_0 K_{\talpha+1}(kx_0) \right).
\end{equation}

We have now to impose the subsidiary condition \eqref{vec:cond}, which
in terms of the Lorentz index fields reads 
\begin{gather}
\label{vec:cond2}
  x_0 \partial_\mu \tA_\mu - d \tA_0=0.\\
\intertext{Inserting \eqref{vec:a0} and \eqref{vec:ai} into \eqref{vec:cond2}
yields}
\label{vec:a0ai}
  a_0 \left(\talpha -\frac{d}{2}+1\right) = i a_i k_i,
\end{gather}
which determines $a_0$ in the generic case of massive vector fields, but leaves
it undetermined in the massless case. In order to find a prescription,
which is valid for both cases, let us first impose the boundary
conditions on the fields $\tA_i$. It is useful to write
\begin{equation}
\label{vec:abb}
  a_i=b_i+bk_i.
\end{equation}
Setting $x_0=\epsilon$ in \eqref{vec:ai} we then find
\begin{equation}
\label{vec:dirich}
  b_i K_\talpha + k_i \left[ b K_\talpha +ia_0 \frac{\epsilon}{k} 
  K_{\talpha+1} \right] = 
  \epsilon^{-\frac{d}{2}} \tA_{\epsilon,i}(\bvec{k}), 
\end{equation}
where the argument $k\epsilon$ of the modified Bessel functions has been
omitted and $\tA_{\epsilon,i}(\bvec{k})$ denotes the Fourier transform
of the Dirichlet boundary value of the field $\tA_i$. We can determine
$b_i$ and $a_0$ from \eqref{vec:dirich} by identifying the first term on 
the l.h.s.\ with the r.h.s.\ and demanding that the second term on the 
l.h.s.\ is zero. This yields
\begin{align}
\label{vec:bi}
  b_i &= \epsilon^{-\frac{d}{2}}
  \frac{\tA_{\epsilon,i}(\bvec{k})}{K_\talpha},\\
\label{vec:a0det}
  a_0 &= i \frac{kbK_\talpha}{\epsilon K_{\talpha+1}}.
\end{align}
Substituting \eqref{vec:abb} and \eqref{vec:a0det} into \eqref{vec:a0ai}
we find the missing coefficcient 
\begin{equation}
\label{vec:b}
  b = \frac{b_i k_i}{k^2} \frac{k\epsilon K_{\talpha+1}}{(1-\tDelta)
  K_\talpha - k\epsilon K_{\talpha-1}},
\end{equation}
where a functional relation of the modified Bessel functions has been
used to rearrange the denominator and we have defined
$\tDelta= \talpha +d/2$.

Let us use the AdS/CFT correspondence to calculate the 
two-point functions of currents $J_i$, which couple to the massive vector 
fields $A_{0,i}$. After integration by parts and using \eqref{vec:eqnmot} 
the action \eqref{vec:action} takes the value
\begin{equation}
\label{vec:action2}
  I = -\frac12 \int d^dx\, \epsilon^{-d} \tA_{\epsilon,i} \left[-\tA_{\epsilon,i}
  +\epsilon \tilde F_{\epsilon,0i} \right],
\end{equation}
where $\tilde F_{0i} = \partial_0 \tA_i - \partial_i \tA_0$ contains 
the interesting part. Using the solutions \eqref{vec:a0} and \eqref{vec:ai} 
with the coefficcients obtained in \eqref{vec:abb}, \eqref{vec:bi}, 
\eqref{vec:a0det} and \eqref{vec:b} one finds
\begin{multline}
\label{vec:f0i}
  \tilde F_{\epsilon,0i} = \left( \frac{d}{2}-\talpha \right) 
  \frac1\epsilon \tA_{\epsilon,i} \\
  + \kint \eikx
  \tA_{\epsilon,j}(\bvec{k}) k \frac{K_{\talpha-1}}{K_\talpha} 
  \left[-\delta_{ij} + \frac{k_ik_j}{k^2} 
  \frac{k\epsilon K_{\talpha+1}}{(\tDelta-1) K_\talpha 
  + k\epsilon K_{\talpha-1}} \right].
\end{multline}
We take the limit $\epsilon\to0$ by substituting the first
terms of the series expansion of the modified Bessel functions in
\eqref{vec:f0i}. The series expansion is given by 
\begin{equation}
\label{vec:bessel}
  K_\nu(z) = z^{-\nu} 2^{\nu-1} \Gamma(\nu) \left[ 1-
  \left(\frac{z}{2}\right)^{2\nu} \frac{\Gamma(1-\nu)}{\Gamma(1+\nu)}
  +\cdots \right],
\end{equation}
where the dots indicate terms of order $z^{2n}$ and $z^{2\nu+2n}$ 
($n=1,2,\ldots$). Our experience from the scalar field \cite{Mueck1}
tells us that the relevant terms are proportional to 
$k^{2\talpha}\delta_{ij}$ and $k^{2\talpha-2} k_i k_j$. We obtain these
by keeping only the leading order terms for the denominators in
\eqref{vec:f0i} and using the appropriate terms for the numerators. In
particular, the term $k^{2\talpha}$ from \eqref{vec:bessel} is needed
only for $K_{\talpha-1}$ in the numerator of \eqref{vec:f0i}. One
obtains
\begin{multline}
\label{vec:f0iapp}
  \tilde F_{\epsilon,0i} = \left( \frac{d}{2}-\talpha \right) 
  \frac1\epsilon \tA_{\epsilon,i} + \left(\frac\epsilon2 \right)^{2\talpha-1}
  \frac{\Gamma(1-\talpha)}{\Gamma(\talpha)} \\
  \times \kint \eikx
  \tA_{\epsilon,j}(\bvec{k}) \left(-k^{2\talpha} \delta_{ij} +
  \frac{2\talpha}{\tDelta-1} k^{2\talpha-2} k_i k_j +\cdots\right),
\end{multline}
where the dots denote all other terms representing either contact
terms in the two-point function or terms of higer order in $\epsilon$. 
Performing the integrals in \eqref{vec:f0iapp} and inserting the result
into \eqref{vec:action2} yields
\begin{multline}
\label{vec:action3} 
  I= \frac12 \left(\talpha +1 -\frac{d}{2} \right) \int d^dx\, \epsilon^{-d} 
  \tA_{\epsilon,i}(\bvec{x}) \tA_{\epsilon,i}(\bvec{x}) \\
  - \frac12 \frac{2\tc\talpha\tDelta}{\tDelta-1} 
  \int d^dx d^dy \tA_{\epsilon,i}(\bvec{x})\tA_{\epsilon,i}(\bvec{y})
  \frac{\epsilon^{2(\tDelta-d)}}{|\bvec{x}-\bvec{y}|^{2\tDelta}} 
  \left(\delta_{ij} - 2
  \frac{(x-y)_i(x-y)_j}{|\bvec{x}-\bvec{y}|^2} \right)+\cdots,
\end{multline}
with
\[ \tc = \frac{\Gamma(\tDelta)}{\pi^\frac{d}{2}
  \Gamma(\talpha)}. \]
Identifying 
\begin{equation}
\label{vec:limit}
  A_{0,i} (\bvec{x}) = \lim_{\epsilon\to0} \epsilon^{\tDelta-d}
  \tA_{\epsilon,i}(\bvec{x})
\end{equation}
and using the AdS/CFT correspondence of the form
\begin{equation}
\label{vec:adscft}
  \exp(-I_{AdS}) \equiv \left\langle \exp\left(\int d^dx \,
  J_j(\bvec{x}) A_{0,j}(\bvec{x}) \right) \right\rangle,
\end{equation}
we can read off from \eqref{vec:action3} the finite distance two-point 
function as
\begin{equation}
\label{vec:2point}
  \left\langle J_i(\bvec{x}) J_j(\bvec{y}) \right\rangle = 
  \frac{2\tc\talpha\tDelta}{\tDelta-1}
  \left(\delta_{ij} - 2
  \frac{(x-y)_i(x-y)_j}{|\bvec{x}-\bvec{y}|^2} \right) 
  |\bvec{x}-\bvec{y}|^{-2\tDelta},
\end{equation}
which is of the form dictated by conformal invariance. It shows in
particular that $J_i$ has the conformal dimension $\tDelta$.
This is of course as expected, but in view of the fact that the integrals 
in \eqref{vec:f0iapp} have to combine to give exactly the terms in
parenthesis in \eqref{vec:2point} it is a non-trivial check of our
derivation. Moreover, our result coincides for the massless case with the 
one obtained in \cite{Freedman}.  

In contrast to the two-point function, which is determined by a boundary 
integral, interactions involve integrals over the volume of AdS$_{d+1}$.
Hence, higher correlation functions are not sensitive to when the limit 
$\epsilon\to0$ is taken and we shall take it for the fields $A_\mu$.
Substituting \eqref{vec:abb}, \eqref{vec:bi}, \eqref{vec:a0det} and
\eqref{vec:b} into \eqref{vec:ai} and replacing $K_\nu(k\epsilon)$ by
the leading order term of its asymptotic expansion \eqref{vec:bessel} 
one finds
\begin{equation}
\label{vec:aibulk}
  A_i^{bulk}(x) =\frac{\tc\tDelta}{\tDelta-1} \int d^dy\,
  A_{0,j}(\bvec{y})
  \frac{x_0^{\tDelta-1}}{\left(x_0^2+|\bvec{x}-\bvec{y}|^2\right)^\tDelta}
  \left(\delta_{ij}-2\frac{(x-y)_i(x-y)_j}{x_0^2+|\bvec{x}-\bvec{y}|^2}\right).
\end{equation}
Similarly, taking the limit in \eqref{vec:a0} yields
\begin{equation}
\label{vec:a0bulk}
  A_0^{bulk}(x) =-\frac{2\tc\tDelta}{\tDelta-1} \int d^dy\,
  A_{0,j}(\bvec{y}) \frac{x_0^\tDelta(x-y)_j}{\left(x_0^2+
  |\bvec{x}-\bvec{y}|^2\right)^{\tDelta+1}}.
\end{equation}

\section{The Free Dirac Field}
\label{spin}
Let us start with the action
\begin{equation}
\label{spin:action}
  I[\bar\psi,\psi] = \int d^{d+1}x \sqrt{g}\, \bar\psi(x) (\slD-m)
  \psi(x) + G \int d^dx \sqrt{h}\, \bar\psi(\bvec{x})\psi(\bvec{x}),
\end{equation}
where we supplemented the dynamical bulk action with a surface term
\cite{Henningson} with an as yet undetermined coefficcient $G$. The
surface term is necessary in order to obtain a two-point function of
spinors in the boundary conformal field theory. The
equation of motion for $\psi$ derived from the action
\eqref{spin:action} is the Dirac equation
\begin{equation} 
\label{spin:psieqn}
  (\slD-m) \psi(x) = \left( x_0 \gamma_\mu \partial_\mu -
  \frac{d}{2} \gamma_0 -m \right) \psi(x) =0,
\end{equation} 
where the matrices $\gamma_\mu$ are the Dirac matrices of $d+1$
dimensional Euclidian space, i.e.\ $\gamma_\mu\gamma_\nu+
\gamma_\nu\gamma_\mu=2\delta_{\mu\nu}$. Acting with
$\gamma_\mu\partial_\mu$ on \eqref{spin:psieqn} one obtains the
second order differential equation 
\begin{equation}
\label{spin:psieqn2}
  \left[ \partial_\mu \partial_\mu-\frac{d}{x_0}\partial_0 -
  \frac{1}{x_0^2} \left( m^2 -\frac{d^2}{4}-
  \frac{d}{2}-\gamma_0m\right)\right] \psi(x) =0.
\end{equation}
The solution of \eqref{spin:psieqn2}, which does not diverge for 
$x_0\to \infty$, is obtained in a similar fashion 
as in the scalar and vector cases and is given by 
\begin{equation}
\label{spin:psi2}
  \psi(x) = \kint \eikx x_0^{\frac{d+1}{2}} \left(a^+(\bvec{k})
  K_{m-\frac12}(kx_0) + a^-(\bvec{k}) K_{m+\frac12}(kx_0) \right),
\end{equation}
where the spinors $a^\pm$ satisfy $\gamma_0a^\pm = \pm a^\pm$. The
expression \eqref{spin:psi2} is in general not a solution of the Dirac 
equation \eqref{spin:psieqn}. In fact, substituting \eqref{spin:psi2} 
into \eqref{spin:psieqn} we find that the spinors $a^+$ and $a^-$ must 
be related by
\begin{equation}
\label{spin:acond}
  a^- = \frac{i}{k} k_i \gamma_i a^+.
\end{equation}

Our next task is to impose boundary conditions on the solution
\eqref{spin:psi2}. However, there is a major difference to the scalar 
and vector cases. The origin of this difference lies in the nature of
the differential equations, which serve as the equations of motion for
the fields. In the scalar case \cite{Mueck1} and vector case
(cf.\ Sec.~\ref{vec}) we have second order differential equations.
Hence, we could impose two sets of boundary data, namely the field and its
derivative. Instead of the latter we demand that the field be well
behaved in the volume of AdS$_{d+1}$, i.e.\ for $x_0\to\infty$, which
yields a unique solution to the Dirichlet problem. On the other
hand, the Dirac equation \eqref{spin:psieqn} is a first order
differential equation. The $x_0\to\infty$ behaviour of the solutions of the 
Dirac equation is crucial from the AdS field theory point of view and cannot 
be abandoned. Hence, only half of the general solutions are available for
fitting the boundary data, which means that only half the components of
the spinor $\psi$ can be prescribed on the boundary, the other half being 
fixed by a relation, which will be determined in a moment. 
This result is important also from a CFT point of view. 
Considering the boundary term of the action
\eqref{spin:action} we realize that, if one could prescribe the entire
boundary spinor, then there would be only a contact term in the CFT
two-point function. The trade-off is that we can obtain only correlators
for spinors, which have half the number of components as the field 
$\psi$. This means that the boundary spinors are Weyl or Dirac spinors 
for $d$ even or odd, respectively \cite{Henningson}. 

Letting $x_0=\epsilon$ in \eqref{spin:psi2} we find 
\begin{equation}
\label{spin:psibcond}
  \psi_\epsilon(\bvec{k}) = \epsilon^{\frac{d+1}{2}} \left( K_{m-\frac12} 
  +i \frac{k_i\gamma_i}{k} K_{m+\frac12} \right) a^+(\bvec{k}), 
\end{equation}
where $\psi_\epsilon(\bvec{k})$ is the Fourier transform of the boundary 
spinor and we have omitted the argument $k\epsilon$ of the modified Bessel 
functions. We can determine $a^+$ from \eqref{spin:psibcond} in two
ways, namely by
\begin{align}
\label{spin:a1}
  a^+(\bvec{k}) &= \epsilon^{-\frac{d+1}{2}}
  \frac{\psi_\epsilon^+(\bvec{k})}{K_{m-\frac12}}\\
\intertext{or}
\label{spin:a2}
  a^+(\bvec{k}) &= \epsilon^{-\frac{d+1}{2}} \frac{k_i\gamma_i}{ik}
  \frac{\psi_\epsilon^-(\bvec{k})}{K_{m+\frac12}}.\\
\intertext{where $\psi^\pm_\epsilon = \frac12 (1\pm\gamma_0)
\psi_\epsilon$. Substituting \eqref{spin:a2} into \eqref{spin:a1} we
find that $\psi^+_\epsilon$ and $\psi^-_\epsilon$ are related by}
\label{spin:pmrel}
  \psi_\epsilon^+(\bvec{k}) &= -i \frac{k_i\gamma_i}{k} 
  \frac{K_{m-\frac12}}{K_{m+\frac12}} \psi_\epsilon^-(\bvec{k}).
\end{align}
The question as to which of the functions $\psi_\epsilon^\pm$ should be
used as boundary data is, in general, not a matter of choice, but 
is dictated by the $\epsilon\to0$ limit. Here we have to distinguish 
three cases. If $m>0$, $K_{m-\frac12}$ diverges slower than 
$K_{m+\frac12}$ for $\epsilon\to0$ and thus we find that 
$\psi_\epsilon^+\to0$, if we fix 
$\psi_\epsilon^-$. This is in agreement with the condition found in 
\cite{Henningson}. On the other hand, we cannot prescribe 
$\psi_\epsilon^+$ for $m>0$, as $\psi_\epsilon^-$ would then diverge. 
The case $m<0$ is just vice versa. For $m=0$ we have
$K_{-\frac12}=K_{\frac12}$ and hence one can prescribe either of the
functions $\psi_\epsilon^\pm$. 

We shall in the following consider the case $m\ge0$. Inserting 
\eqref{spin:a2} and \eqref{spin:acond} into \eqref{spin:psi2}
we finally find 
\begin{equation}
\label{spin:psi}
  \psi(x) = \kint \eikx \left(\frac{x_0}{\epsilon}\right)^\frac{d+1}{2}
  \left(-i\frac{k_i\gamma_i}{k} K_{m-\frac12}(kx_0) +
  K_{m+\frac12}(kx_0) \right)
  \frac{\psi_\epsilon^-(\bvec{k})}{K_{m+\frac12}(k\epsilon)}.
\end{equation}
In a similar fashion one can solve the equation of motion for the
conjugate spinor,
\begin{equation}
\label{spin:psibeqn}
  \bar\psi(x)(\overset{\leftarrow}{\slD}+m) = \bar\psi(x)\left( 
  \overset{\leftarrow}{\partial}_\mu \gamma_\mu x_0
  -\frac{d}{2} \gamma_0 +m \right) = 0.
\end{equation}
The solution in the case $m\ge0$ is 
\begin{equation}
\label{spin:psib}
  \bar\psi(x) = \kint \eikx 
  \left(\frac{x_0}{\epsilon}\right)^\frac{d+1}{2} 
  \frac{\bar\psi_\epsilon^+(\bvec{k})}{K_{m+\frac12}(k\epsilon)}
  \left(i\frac{k_i\gamma_i}{k} K_{m-\frac12}(kx_0) +
  K_{m+\frac12}(kx_0) \right),
\end{equation}
where $\bar\psi_\epsilon^\pm=\bar\psi_\epsilon \frac12(1\pm\gamma_0)$. 
Again we find a relation between the components of the boundary spinor, 
which is given by
\begin{equation}
\label{spin:pmbrel}
  \bar\psi_\epsilon^-(\bvec{k}) = \bar\psi_\epsilon^+(\bvec{k}) 
  i \frac{k_i\gamma_i}{k} \frac{K_{m-\frac12}}{K_{m+\frac12}}.
\end{equation}

Let us turn now to the two-point function for the boundary spinors 
$\chi^+$ and $\bar\chi^-$, which couple to 
$\bar\psi_0^+$ and $\psi_0^-$, respectively. Inserting the solutions of
the equations of motion into the action \eqref{spin:action}, the bulk
term vanishes and the surface term can be written as 
\begin{equation}
\label{spin:actionsplit}
  I=G \epsilon^{-d} \kint \left(\bar\psi^+(\bvec{k})\psi^+(-\bvec{k}) 
  + \bar\psi^-(\bvec{k})\psi^-(-\bvec{k}) \right).
\end{equation}
Using the relations \eqref{spin:pmrel} and \eqref{spin:pmbrel} one finds
\begin{equation}
\label{spin:action2}
  I=G\epsilon^{-d}\int d^dx d^dy \kint
  \e^{i\bvec{k}\cdot(\bvec{x}-\bvec{y})}
  \bar\psi_\epsilon^+(\bvec{x})\left( 2i
  \frac{k_i\gamma_i}{k} \frac{K_{m-\frac12}}{K_{m+\frac12}} \right)
  \psi_\epsilon^-(\bvec{y}).
\end{equation}
We use the expansion \eqref{vec:bessel} for the modified Bessel functions
in the numerator and the leading order term in the denominator. Hence, 
we find after integration
\begin{equation}
\label{spin:action3}
  I= - 2 \hat c G \int d^dx d^dy\, \bar\psi_0^+(\bvec{x})
  \frac{\gamma_i(x_i-y_i)}{|\bvec{x}-\bvec{y}|^{d+2m+1}}
  \psi_0^-(\bvec{y}),
\end{equation}
where we defined
\begin{equation}
\label{spin:psi0}
  \psi_0^- = \lim_{\epsilon\to0} \epsilon^{m-\frac{d}{2}} \psi_\epsilon^-
  \quad \text{and} \quad \bar\psi_0^+= 
  \lim_{\epsilon\to0} \epsilon^{m-\frac{d}{2}} \bar\psi_\epsilon^+
\end{equation}
for the $\epsilon\to0$ limit and
\[ \hat c =
\frac{\Gamma\left(\frac{d+1}{2}+m\right)}{\pi^\frac{d}{2}\Gamma
\left(m+\frac12\right)}. \]
In the case $m=0$ the $k$ integral in \eqref{spin:action2} can be done
without the asymptotic expansion and leads to the same result. Using the
AdS/CFT correspondence 
\begin{equation}
\label{spin:adscft}
  \exp(-I_{AdS}) \equiv \left\langle \exp\left(\int d^dx \,(
  \bar\chi^-\psi_0^- + \bar\psi_0^+\chi^+) \right) \right\rangle,
\end{equation}
the two-point function reads
\begin{equation}
\label{spin:2point}
  \langle \chi^+(\bvec{x})\bar\chi^-(\bvec{y})\rangle = 2 \hat c G 
  \frac{\gamma_i(x_i-y_i)}{|\bvec{x}-\bvec{y}|^{d+2m+1}}.
\end{equation}
Hence, the spinors $\chi$ and $\bar\chi$ have the conformal dimension
$m+\frac{d}{2}$. Our result agrees up to the appropriate normalization 
with the one found in \cite{Henningson}. 

For calculating interactions we are interested in the bulk behaviour of
the spinors $\psi$ and $\bar\psi$. It is obtained by replacing
$K_{m+\frac12}(k\epsilon)$ by the leading order term of its asymptotic
expansion in \eqref{spin:psi} and \eqref{spin:psib}.
One finds the expressions
\begin{align}
\label{spin:psibulk}
  \psi^{bulk}(x) &= \hat c \int d^dy\, 
  \left[x_0 - \gamma_i (x_i-y_i)\right] \left( x_0^2 +
  |\bvec{x}-\bvec{y}|^2\right)^{-\frac{d+1}{2}-m} x_0^{\frac{d}{2}+m}
  \psi_0^-(\bvec{y})\\
\intertext{and}
\label{spin:psibbulk}
  \bar\psi^{bulk}(x) &= \hat c \int d^dy\, \bar\psi_0^+(\bvec{y})
  \left[x_0 + \gamma_i (x_i-y_i)\right] \left( x_0^2 +
  |\bvec{x}-\bvec{y}|^2\right)^{-\frac{d+1}{2}-m} x_0^{\frac{d}{2}+m},
\end{align}
which coincide with those in \cite{Henningson} up to normalization. 
A good check of the
derivation of these expressions
is provided by the case $m=0$. Since $K_{\pm\frac12}(z)=
\sqrt{\frac{\pi}{2z}} \e^{-z}$, it is possible to carry out the 
integral in \eqref{spin:psi} with the result
\begin{equation}
\label{spin:psim0}
  \psi(x)= \int d^dy\,
  \frac{\Gamma\left(\frac{d+1}{2}\right)}{\pi^\frac{d}{2}\Gamma
  \left(\frac12\right)}x_0^\frac{d}{2} \left[(x_0-\epsilon)^2 +
  |\bvec{x}-\bvec{y}|^2\right]^{-\frac{d+1}{2}}
  \left[x_0-\epsilon-\gamma_i(x_i-y_i)\right]\psi_0^-(\bvec{y}).
\end{equation}
  
\section{Interaction between Spinor and Gauge Fields}
\label{int}
Calculating the first order interaction between the spinor and massless
vector fields serves two purposes. First, it provides another detail of
the AdS/CFT correspondence in form of the vector-spinor-spinor 
three-point function. In contrast to the scalar three-point function, 
conformal symmetry does not fix, but only restricts
the form of this particular three-point function \cite{Osborn}. 
Hence, the calculation
will yield more than just a coefficcient in front of a universal
function. Second, a check of the Ward identity corresponding to the 
gauge invariance will reveal that no 
supplementary surface term of the order of the gauge coupling is needed.

We shall use the action for minimally coupled spinor and gauge fields,
together with the spinor surface term,
\begin{equation}
\label{int:action}
  I = \int d^{d+1}x \sqrt{g} \left[ \frac14 F_{\mu\nu} F^{\mu\nu} 
  + \bar\psi \left(\slD-iq \slA -m\right) 
  \psi \right]+ G \int d^dx \sqrt{h}\, \bar\psi\psi. 
\end{equation}
The equations of motion derived from \eqref{int:action} are 
\begin{align}
\label{int:feqn}
  \nabla_\mu F^{\mu\nu} &= -iq \e^\nu_a\bar\psi\gamma_a\psi,\\
\label{int:psieqn}
  (\slD-m)\psi &= iq \slA \psi \\
\intertext{and its conjugate}
\label{int:psibeqn}
  \bar\psi(\overset{\leftarrow}{\slD}+m) &= -iq \bar\psi \slA.
\end{align}
We split the gauge field into its free part $A^{(0)}$ and the remainder 
$A^{(1)}$. Substituting \eqref{int:psieqn} into \eqref{int:action} and
using the equation of motion for $F^{(0)}$, we find
\begin{equation}
\label{int:action2}
  I= \int d^dx \epsilon^{-d} \left(-\frac1{2\epsilon} A^{(0)}_i F^{(0),0i}
  -\frac1{\epsilon} A^{(1)}_i F^{(0),0i} + G \bar\psi\psi \right) + \Ord(q^2).
\end{equation}
Most importantly, the bulk terms vanish! Moreover, using the appropriate 
Green's function to calculate $A^{(1)}$ (cf.\ \cite{Mueck1} for the
scalar field analogue), we realize that also the second term in
\eqref{int:action2} is zero. The first term only yields the two-point function
for the conserved currents $J$. However, the last term will give the
two-point function for the spinors and the three-point function coupling $J$ 
and the spinors. This surprising fact comes about as follows. Going back to the 
derivation of the spinor two-point function, we realize that it was
generated by the relations \eqref{spin:pmrel} and \eqref{spin:pmbrel}
between the $+$ and $-$ components of the spinors on the boundary. These
relations will be altered by the presence of the interaction. Writing
\begin{align}
\label{int:psi}
  \psi(x) &= \psi^{(0)}(x) + \psi^{(1)}(x) + \Ord(q^2),\\
\label{int:psi1}
  \psi^{(1)}(x) &= iq \int d^{d+1}y \sqrt{g}\, S(x,y) \slA(y)
  \psi^{(0)}(y),
\end{align}
where $S(x,y)$ is the spinor Green's function defined by
\begin{equation}
\label{int:S}
  (\slD_x-m)S(x,y) = \frac{\delta(x-y)}{\sqrt{g(x)}},
\end{equation}
we find using \eqref{spin:pmrel}
\begin{equation}
\label{int:pmrel}
  \psi^+(\bvec{k}) = -i \frac{k_i\gamma_i}{k}
  \frac{K_{m-\frac12}}{K_{m+\frac12}} \psi^-(\bvec{k}) + 
  \frac{1+\gamma_0}{2} \left(1+i\frac{k_i\gamma_i}{k}
  \frac{K_{m-\frac12}}{K_{m+\frac12}} \right) \psi^{(1)}(\bvec{k}) + \Ord(q^2),
\end{equation}
where we omitted the argument $k\epsilon$ of the modified Bessel
functions. 
Similarly, one finds for the conjugate field 
\begin{align}
\label{int:psib}
  \bar\psi(x) &= \bar\psi^{(0)}(x) + \bar\psi^{(1)}(x) + \Ord(q^2),\\
\label{int:psib1}
  \bar\psi^{(1)}(x) &= iq \int d^{d+1}y \sqrt{g}\, \bar\psi^{(0)}(y) 
  \slA(y) \bar S(y,x),
\end{align}
with\footnote{The relation between $\bar S$ and $S$ is of no importance
here.}
\begin{equation}
\label{int:Sb}
  \bar S(y,x) (\overset{\leftarrow}{\slD}+m) = 
  -\frac{\delta(x-y)}{\sqrt{g(x)}},
\end{equation}
and, using \eqref{spin:pmbrel}
\begin{equation}
\label{int:pmbrel}
  \bar\psi^-(\bvec{k}) = \bar\psi^+(\bvec{k}) i\frac{k_i\gamma_i}{k}
  \frac{K_{m-\frac12}}{K_{m+\frac12}} + 
  \bar\psi^{(1)}(\bvec{k}) \left(1-i\frac{k_i\gamma_i}{k}
  \frac{K_{m-\frac12}}{K_{m+\frac12}} \right) \frac{1-\gamma_0}{2}  +
  \Ord(q^2).
\end{equation}
Substituting \eqref{int:pmrel} and \eqref{int:pmbrel} into the spinor
surface term in the form \eqref{spin:actionsplit}, one finds that the
contribution to the action of first order in $q$ is 
\begin{multline}
\label{int:action3}
  I^{(1)}= G\epsilon^{-d} \kint \left[ \left(\bar\psi^{(0)+}(\bvec{k}) - 
  \bar\psi^{(0)-}(\bvec{k}) \right) \psi^{(1)}(-\bvec{k}) \right.\\ -
  \left.\bar\psi^{(1)}(-\bvec{k}) \left( \psi^{(0)+}(\bvec{k}) - 
  \psi^{(0)-}(\bvec{k}) \right) \right] +\Ord(q^2).
\end{multline}

On the other hand, from \eqref{int:Sb} and \eqref{int:S} one can obtain
\begin{align}
\label{int:psi0}
  \psi^{(0)}(x) &= \epsilon^{-d} \int d^dy\, \bar S(x,\bvec{y})
  \left( \psi^{(0)+}(\bvec{y}) - \psi^{(0)-}(\bvec{y}) \right)\\
\intertext{and}
\label{int:psib0}
  \bar\psi^{(0)}(x) &= -\epsilon^{-d} \int d^dy\, 
  \left(\bar\psi^{(0)+}(\bvec{y}) - \bar\psi^{(0)-}(\bvec{y}) \right)
  S(\bvec{y},x),
\end{align}
respectively. Inserting \eqref{int:psi1}, \eqref{int:psib0},
\eqref{int:psib1} and \eqref{int:psi0} into \eqref{int:action3}
one then finds 
\begin{equation}
\label{int:action4}
  I^{(1)} = -2Giq \int d^{d+1}x \sqrt{g}\, \bar\psi^{(0)} \slA
  \psi^{(0)}.
\end{equation}
Eqn.\ \eqref{int:action4} has the same form as the minimal coupling
term, but is multiplied by $2G$. It is determined by
a bulk integral, which means that the bulk behaviour for the fields can
be used. Substituting
\eqref{vec:aibulk}, \eqref{vec:a0bulk} (with $\tDelta=d-1$), 
\eqref{spin:psibulk} and
\eqref{spin:psibbulk} into \eqref{int:action4}, the following tedious 
calculation involves Feynman parameterization of the denominator and
heavy numerator algebra. The result is 
\begin{multline}
\label{int:3point}
  \langle J_j(\bvec{x}_2) \chi^+(\bvec{x}_1) \bar\chi^-(\bvec{x}_3) \rangle 
  = \frac{-iGq\hat c \Gamma\left(\frac{d}{2}\right)}{
  \pi^\frac{d}{2} (d-1+2m)}\\ \times 
  \left[ (d-2) \gamma_i\gamma_j\gamma_k \frac{x_{12i} x_{23k}}{x_{12}^d
  x_{23}^d x_{13}^{2m+1}} +(2m+1) \gamma_i x_{13i}
  \frac{x_{12}^2x_{23j}+
  x_{23}^2x_{12j}}{x_{12}^dx_{23}^dx_{13}^{2m+3}} \right],
\end{multline}
where $\bvec{x}_{ab}=\bvec{x}_a-\bvec{x}_b$.
After further algebra one finds that \eqref{int:3point} can be
written in the form
\begin{multline}
\label{int:3point2}
  \langle J_j(\bvec{x}_2) \chi^+(\bvec{x}_1) \bar\chi^-(\bvec{x}_3) \rangle 
  = \frac{-iGq\hat c \Gamma\left(\frac{d}{2}\right)}{
  \pi^\frac{d}{2} (d-1+2m)}
  \frac{1}{x_{12}^d x_{23}^{d-2} x_{13}^{2m}}
  \frac{\gamma_i x_{13i}}{x_{13}}\\ \times
  \left(\delta_{jk}-2\frac{x_{23j}x_{23k}}{x_{23}^2} \right)
  \left(\frac{x_{13l}}{x_{13}^2}-\frac{x_{23l}}{x_{23}^2} \right)
  \left[(d-2)\gamma_l\gamma_k+(2m+1)\delta_{kl}\right],
\end{multline}
which is a specific case of the general expression dictated 
by conformal invariance \cite{Osborn}.

Finally, let us confirm the Ward identity \cite{Francesco}
\begin{equation}
\label{int:ward}
  \partdiff{x_2^j} 
  \langle J_j(\bvec{x}_2) \chi^+(\bvec{x}_1) \bar\chi^-(\bvec{x}_3) \rangle
  = -iq \langle\chi^+(\bvec{x}_1)\bar\chi^-(\bvec{x}_3)\rangle
  [\delta(\bvec{x}_{23})-\delta(\bvec{x}_{12})].
\end{equation}
From \eqref{int:3point} one finds 
\begin{equation}
\label{int:ward2}
  \partdiff{x_2^j} 
  \langle J_j(\bvec{x}_2) \chi^+(\bvec{x}_1) \bar\chi^-(\bvec{x}_3) \rangle
  = -iGq 2\hat c \frac{\gamma_i x_{13i}}{x_{13}^{d+2m+1}} 
  [\delta(\bvec{x}_{23})-\delta(\bvec{x}_{12})].
\end{equation}
Comparing \eqref{int:ward2} and \eqref{spin:2point} with \eqref{int:ward}
we see that the Ward identity is satisfied. This result is significant,
since is tells us that, to first order in $q$, no supplementary surface 
term except the one used already for the free Dirac field is required 
in the action for interacting fields.

\section{Conclusions}
\label{conc}
In the present paper we used the AdS/CFT correspondence to calculate CFT
correlators from the classical AdS theories of vector and Dirac fields. We 
took care to address the proper treatment of the $\epsilon\to0$ limit
when calculating the two-point functions. 
As for the scalar field \cite{Mueck1,Freedman}, this was particularly 
important for the vector field with non zero mass. 

Our calculation for the free Dirac field revealed the full details as to why 
only half the number of spinor components
can be given as boundary data. For odd $d$ this is exactly what one wants, 
because the boundary spinor representation has only half the number of
components as the bulk spinors. For even $d$ the dimensions of the
spinor representations are the same and $\gamma_0$ acts as the chirality
operator on the boundary spinors. This means that for even $d$ we calculated
only the correlation functions for chiral spinors. However, the
formalism can be extended to Dirac spinors by coupling $\chi^-$ to an
AdS spinor $\psi_1$ with positive mass $m$ and $\chi^+$ to a field $\psi_2$ 
with mass $-m$. 

Minimally coupling the Dirac and massless vector field, we calculated
the CFT vector-spinor-spinor three-point function. The result should be
interesting from a CFT point of view, as the form of this correlator is
not totally fixed by conformal invariance. Thus, our result could  
indicate, which CFT is obtained by the AdS/CFT correspondence.
Finally, we confirmed the validity of the Ward identity and found that
no interaction surface terms are required in the action.

\section*{Acknowledgements}
We would like to thank D.~Freedman and K.~Sfetsos for interesting 
correspondence.

This work was supported in part by an operating grant from NSERC.
W.~M.\ gratefully acknowledges the support with a Graduate Fellowship
from Simon Fraser University.

\renewcommand{\baselinestretch}{1}

\end{document}